\documentstyle[12pt]{article}

\begin{document}

\noindent {\large {\bf Inverse momentum expectation values for hydrogenic systems}}

\vspace{1in}

\noindent {\em R. Delbourgo and D. Elliott}

\medskip

{\small \noindent School of Mathematics and Physics, University of Tasmania\\
Private Bag 37, GPO, Hobart, Australia 7001 \vspace{1in} }

{\small \noindent {\em Abstract} }

{\small \noindent By using the Fourier transforms of the general hydrogenic 
bound state wave functions (as ultraspherical polynomials) one may find 
expectation values of arbitrary functions of momentum $p$. In this manner the 
effect of a reciprocity perturbation $b/p$ can be evaluated for all hydrogenic 
states.}

\medskip

\section{Motivation}
Many years ago and long before the elementary particle spectrum was thoroughly
explored experimentally, Born and Green \cite{BG} proposed the principle of 
reciprocity in an attempt to determine the mass spectrum of fermions and bosons. 
The principle flowed naturally from observed covariance of equations of motion 
under the momentum-position substitution rule $(P,R)\rightarrow(-bR,P/b)$, 
with an appropriate scale $b$, but it went a lot further in postulating that the 
Hamiltonian was actually {\em invariant} under such a transformation. This was all 
done in a relativistic framework but it soon became apparent that it failed rather
miserably to reproduce the mass spectrum, since all states were essentially
harmonic overtones of a fundamental frequency; thus the idea was soon consigned 
to the dustbin of history. Recently there has been a resurgence in investigating
the concept, not only for its elegance but for its group theoretical features
which incorporate the concept of a maximum force. Much progress has been made
along such lines \cite{Low} but it remains true that the relativistic scheme
is beset with tachyonic state problems if the spectrum is treated along the
lines of Wigner's approach of induced representations, ensuing from the
larger `quaplectic' group.

Even at the classical nonrelativistic level where the idea should leave an
imprint, the consequence for the (undamped) harmonic oscillator is that all 
frequencies are universal, which is patently absurd, and seems to indicate that
the reciprocity concept has no future. However with the realization that damping
with an appropriately small scale $b$ avoids this absurdity, it is worth
following through the idea, at least non-relativistically, for some familiar
potentials and in particular for Coulomb-like ones about which so much is known.
Now nonrelativistic systems placed in a $1/r$ potential with Hamiltonian
\begin{equation}
H(P,R) = P^2/2m - \alpha/R
\end{equation}
have been thoroughly studied over many years and the results are found in
standard textbooks (see eg \cite{Messiah}) of classical and quantum mechanics. 
More complicated systems or small modifications, including certain relativistic 
corrections, can be treated by perturbation theory. If we attempt to make such a
system reciprocity-invariant the Hamiltonian (1) is accompanied by extra terms 
$b^2R^2/2m -\alpha b/P$. Here $b$ is a tiny scale factor which has hardly any 
effect on atomic physics but can influence phenomena on cosmic scales \cite{DL}. 
Because $b$ is so small the main effect of the reciprocity change lies in
the $b/P$ perturbation, so our aim in this paper is to determine the energy level
change on {\em all}\, hydrogenic bound state levels due to it, and not just on the
ground state as was recently done in \cite{DL}. This is the motivation for this 
article apart from its intrinsic mathematical interest.

The neatest way to obtain the expectation value of any function of momentum $f(P)$ is
to evaluate the Fourier transform $\phi(\mathbf p)$ of the hydrogenic wave functions
$\psi({\mathbf x})$ and work out 
\begin{equation}
\langle f(P)\rangle=\int\frac{d^3p}{h^3}\,\phi^*({\mathbf p})f(p)\phi({\mathbf p})
\end{equation}
in the usual way. If $f$ were purely a polynomial it would not be necessary to go
through this process but simply work directly in coordinate space,
\[
\langle f(P)\rangle=\int d^3x\,\psi^*({\mathbf x})f(-i\hbar\nabla)\psi({\mathbf x}),
\] 
because the spatial wave functions $\psi$ are very well known. However, for 
expectation values of the reciprocity term where we are dealing with an inverse
momentum, it is safest to take route (2). Nevertheless the momentum-space 
procedure (2) requires us to evaluate first the Fourier transforms,
\[
\phi({\mathbf p})=\int d^3x\,{\rm e}^{-i{\mathbf p}.{\mathbf x}/\hbar}\psi({\mathbf x}),
\]
and this is technically tricky! It was first carried out in a classic paper by Podolsky 
and Pauling \cite{PP} but our treatment is slightly different and done in the next 
section. It also differs radically from Hylleraas' differential method \cite{H}.
The required expectation value of $1/P$, the meat of the paper, is given in the 
following section; the answer is nontrivial. Interesting sum rules are worked out next 
and asymptotic estimates conclude the paper.

\section{Momentum wave functions}
We begin by quoting the bound state wave functions for Hamiltonian (1), as stated
in the standard texts \cite{Messiah}:
\begin{equation}
\psi_{n\ell m}({\mathbf x})\equiv \langle r\theta\phi|n\ell m\rangle =
 2\kappa^{3/2}\sqrt{\frac{(n\!-\!\ell\!-\!1)!}{n(n+\ell)!}}{\rm e}^{-\kappa r}
 (2\kappa r)^\ell L_{n-\ell-1}^{2\ell+1}(2\kappa r)\,Y_{\ell m}(\theta,\phi),
\end{equation}
where the quantum numbers $n,\ell,m$ are integers obeying $n >\ell\geq |m|\geq 0$,
with $\kappa\equiv 1/na = m\alpha/n\hbar^2$,
so $a$ connotes the Bohr radius and 
$Y_{\ell m}$ are the spherical harmonics, orthonormal over the unit sphere. The 
associated Laguerre polynomials in (7) obey the orthogonality relations 
(see \cite{Magnus}, \S 5.5.2)
\begin{equation}
\int_0^\infty {\rm e}^{-t}\,t^\nu L_n^\nu(t) L_{n'}^\nu(t)\, dt = 
 \delta_{nn'}\Gamma(n+\nu+1)/\Gamma(n+1).
\end{equation}
Naturally, the wave functions (3) are orthonormal over space,
\[
\int d^3x\,\psi^*_{n\ell m}({\mathbf x})\psi_{n'\ell' m'}({\mathbf x)}=
\delta_{nn'}\delta_{\ell\ell'}\delta_{mm'},
\]
but this is difficult to demonstrate directly from (4) because of the 
different radial weight; in that respect the intrinsic $n$-dependence of
$\kappa$ becomes significant, as shown by Dunkl \cite{Dunkl}.

To be able to obtain the momentum-space wave function, 
\begin{equation}
\phi_{n\ell m}({\mathbf k}) \equiv \langle k \theta_k\phi_k|n\ell m\rangle =
\int d^3x \,{\rm e}^{-i{\mathbf k}.{\mathbf x}} \psi_{n\ell m}({\mathbf x}),
\end{equation}
we make use of the expansion of plane waves into spherical ones (see \cite{Messiah},
eq (B.105)):
\begin{equation}
{\rm e}^{-i{\mathbf k}.{\mathbf x}}=4\pi\sum_{\ell,m}(-i)^\ell j_\ell(kr)
 Y^*_{\ell m}(\theta,\phi)Y_{\ell m}(\theta_k,\phi_k);\quad
j_\ell(z)\equiv \sqrt{\frac{\pi}{2z}}J_{\ell+1/2}(z).
\end{equation}
By this means we arrive at $\phi_{n\ell m}(k,\theta_k,\phi_k)\equiv
(-i)^\ell{\cal P}_{n\ell}(k) Y_{\ell m}(\theta_k,\phi_k)$ and the radial 
momentum wave function
\begin{equation}
{\cal P}_{n\ell}(k)=4\pi\kappa^{3/2}\sqrt{\frac{(n\!-\!\ell\!-\!1)!}{n(n+\ell)!}}
\!\!\int_0^\infty\!\!\!\!\sqrt{\frac{\pi}{2kr}} J_{\ell+1/2}(kr)(2\kappa r)^\ell
{\rm e}^{-\kappa r}L_{n-\ell-1}^{2\ell+1}(2\kappa r)\,r^2 dr.
\end{equation}
This is a formidable integral in general, except for simple cases like $\ell = 0$ or
$\ell = n-1$.

To proceed further, we make use of an integral which {\em can} be found in the 
standard texts (see \cite{Magnus}, \S 3.8.3),
namely
\begin{equation}
\int_0^\infty t^{\nu+1}{\rm e}^{-\beta t}J_\nu(\gamma t)\,dt =
\frac{2^{\nu+1}\beta\gamma^\nu\Gamma(\nu+3/2)}{\sqrt{\pi}(\beta^2+\gamma^2)^{\nu+3/2}},
\end{equation}
and the generating function (see \cite{Magnus}, \S 5.5.2)
\begin{equation}
\sum_{\nu=0}^\infty L_\nu^\alpha(x)\,z^\nu = (1-z)^{-\alpha-1}{\rm e}^{xz/(z-1)}.
\end{equation}
By defining (and later on identifying $\nu = n-\ell-1 \geq 0$) 
\begin{equation}
{\cal P}_\ell(k,z)\equiv \sum_{\nu=0}^\infty \sqrt{\frac{\nu!}
 {(\nu+\ell+1)(\nu+2\ell+1)!\kappa^3}}\,\, {\cal P}_{\nu\ell}(k)\,z^\nu
\end{equation}
we see that such an expansion in powers of $z$ yields the momentum space
hydrogenic functions. This new generating function is a doable integral
like (8),
\begin{eqnarray}
{\cal P}_\ell(k,z)&=&\frac{4\pi^{3/2}(2\kappa)^\ell}{(1-z)^{2\ell+2}\sqrt{2k}}
  \int_0^\infty r^{\ell+3/2}{\rm e}^{-\kappa r(1+z)/(1-z)}J_{\ell+1/2}(kr)\,dr
  \nonumber \\
 &=& \frac{8\pi\kappa\,(4k\kappa)^\ell(1-z^2)(\ell+1)!}
          {(\kappa^2(1+z)^2 + k^2(1-z)^2)^{\ell+2}}.
\end{eqnarray}

Since ultraspherical (Gegenbauer) polynomials arise via the generating series
(see \cite{Magnus}, \S 5.3.2),
\begin{equation}
(1-2xz+z^2)^{-\lambda} = \sum_{\nu=0}^\infty C_\nu^\lambda(x)\,z^\nu;\quad 
 |z|<1,\quad \lambda\neq 0,
\end{equation}
we are led to identify $x=(k^2-\kappa^2)/(k^2+\kappa^2)$, and derive 
\begin{eqnarray}
\frac{(k^2+\kappa^2)^{\ell+2}}{8\pi\kappa\,(4k\kappa)^\ell(\ell+1)!}
{\cal P}_\ell(k,z)&=& (1-z^2)\sum_{\nu=0}^\infty C_\nu^{\ell+2}(x)\,z^\nu=\nonumber\\
\sum_{\nu=0}^\infty [C_\nu^{\ell+2}(x)-C_{\nu-2}^{\ell+2}(x)]\,z^\nu&=&
\sum_{\nu=0}^\infty (\nu+\ell+1)\,C_\nu^{\ell+1}(x)\,z^\nu/(\ell+1),
\end{eqnarray}
on adopting the convention that polynomials of negative degree are 
identically zero and using recurrence relations.
Putting this together with (10), we arrive at the sought-after result for
the Fourier transform, $\phi_{n\ell m}({\mathbf k})=(-i)^\ell
{\cal P}_{n\ell}(k)Y_{\ell m}(\theta_k,\phi_k)$, namely
\begin{equation}
{\cal P}_{n\ell}(k) = 16\pi\kappa^{5/2}\sqrt{\frac{n(n-\ell-1)!}{(n+\ell)!}}
       \frac{(4k\kappa)^\ell\,\ell!}{(k^2+\kappa^2)^{\ell+2}}
       C_{n-\ell-1}^{\ell+1}\left(\frac{k^2-\kappa^2}{k^2+\kappa^2}\right).
\end{equation}
This will form the basis for the forthcoming calculations. [Fock's 
stereographic representation \cite{Fock} is also capable to reproducing (14).]

For the present it only remains to check the normalization of these momentum 
wavefunctions:
\begin{eqnarray}
\int_0^\infty |{\cal P}_{n\ell}(k)|^2\frac{k^2\,dk}{8\pi^3} &=&
 \frac{32 n(n-\ell-1)!(\ell!)^2\kappa^5}{\pi(n+\ell)!}\times\nonumber\\
 & & \,\,\,
 \int_0^\infty\!\!\!\!\frac{(4k\kappa)^{2\ell}}{(k^2+\kappa^2)^{2\ell+4}}\left[
  C_{n-\ell-1}^{\ell+1}\left(\frac{k^2\!-\!\kappa^2}{k^2\!+\!\kappa^2}\right)
 \right]^2k^2dk.
\end{eqnarray}
Changing the integration variable back to $x =(k^2-\kappa^2)/(k^2+\kappa^2)$, 
the rhs of (15) can be simplified to
\[
 \frac{2n(n-\ell-1)!(2^\ell\ell!)^2}{\pi(n+\ell)!}\int_{-1}^1(1-x^2)^{\ell+1/2}
 [C_{n-\ell-1}^{\ell+1}(x)]^2\,(1-x)\,dx  .
\]
Discarding the odd term in $x$ and using orthogonality of the Gegenbauer
polynomials (see \cite{Magnus}, \S 5.3.2),
\[
 \int_{-1}^1C_n^\lambda(x)\,C_{n'}^\lambda(x)\,(1-x^2)^{\lambda-1/2}\,dx=\delta_{nn'} 
 \frac{2^{1-2\lambda}\pi\Gamma(n+2\lambda)}{(\lambda+n)n!(\Gamma(\lambda))^2},
\]
we can satisfy ourselves that the rhs of (15) does indeed reduce to unity; 
so all is well for the next initiative.

\section{Momentum expectation values}
We are ready to tackle the general case,
\[
\langle f(P)\rangle_{n\ell} = \langle n\ell m|f(P)|n\ell m\rangle 
 = \int_0^\infty |{\cal P}_{n\ell}(k)|^2\,f(\hbar k)\frac{k^2\,dk}{8\pi^3}
\]
\begin{equation}
=\frac{2n(n-\ell-1)!(2^\ell\ell!)^2}{\pi(n+\ell)!}\!\!\int_{-1}^1(1-x^2)^{\ell+1/2}
[C_{n-\ell-1}^{\ell+1}(x)]^2\,(1-x)f\left(\hbar\kappa\sqrt{\frac{1+x}{1-x}}\right)\,dx,
\end{equation}
provided of course that the resulting integration is well-behaved so that
$f(P)$ makes sense. As we are going to be dealing with squares of
ultraspherical polynomials, let us substitute $k=\kappa\tan\theta$, making 
$x=-\cos(2\theta)$. It then follows that expression (16) can be recast as
\[
\langle f(P)\rangle_{n\ell}=\frac{4n(n-\ell-1)!(2^\ell\ell!)^2}{\pi(n+\ell)!}
  \int_0^{\pi/2}\!(\sin 2\theta)^{2\ell+2}(1+\cos 2\theta)
  [C_{n-\ell-1}^{\ell+1}(\cos 2\theta)]^2 
\]
\begin{equation}
\qquad\qquad\qquad \times f(\hbar\kappa\tan\theta)\,d\theta.
\end{equation}
In particular, for the inverse momentum we meet the dimensionless integrals
\begin{eqnarray}
\langle \frac{\hbar\kappa}{P}\rangle_{n\ell}&=&
\frac{16n(n\!-\!\ell\!-\!1)!(2^\ell\ell!)^2}{\pi(n+\ell)!}\!\!
  \int_0^{\pi/2}\!\!\!\!(\sin 2\theta)^{2\ell+1}\cos^4\theta\,
  [C_{n-\ell-1}^{\ell+1}(\cos 2\theta)]^2d\theta\nonumber \\
&{\rm or}& \frac{2n(n\!-\!\ell\!-\!1)!(2^\ell\ell!)^2}{\pi(n+\ell)!}\!\!
 \int_{-1}^1 (1-x^2)^\ell[(1+x)C_{n-\ell-1}^{\ell+1}(x)]^2\,dx.
\end{eqnarray}
We shall now show how to evaluate these for special values of $\ell$ before 
handling the most general angular momentum state.

\subsection{The case $\ell=0$}
To demonstrate the nontrivial nature of the problem we firstly turn to the spherical
(S-wave) $\ell=0$ states. Since $C_{n-1}^1(\cos\theta)=\sin(n\theta)/\sin\theta$
(see \cite{Magnus}, \S 5.3.1), we must deal with
\begin{equation}
\langle \frac{\hbar\kappa}{P}\rangle_{n\,0} = \frac{8}{\pi}\int_0^{\pi/2}
  \frac{\sin^2(2n\theta)}{\sin\theta}\cos^3\theta\,d\theta.
\end{equation}
In order to do this (via a recursion procedure) let us define, for integer $\nu\geq 0$,
\begin{equation}
K_\nu(n)\equiv \int_0^{\pi/2}  \frac{\sin^2(2n\theta)}{\sin\theta}
 \cos^{2\nu+1}\theta\,d\theta,
\end{equation}
and treat the case $\nu=0$ first. We have
\begin{equation}
 K_0(n+1)-K_0(n)=\int_0^{\pi/2}\!\!\cot\theta[\sin^2(2n+2)\theta-\sin^22n\theta]
  \,d\theta = 1/(2n+1).
\end{equation}
But in the particular case $n=1$ we have trivially $K_0(1)=1$. It follows from (21) 
that (see \cite{Magnus}, \S 1.2)
\begin{equation}
K_0(n)=\sum_{m=1}^n\frac{1}{2m-1}=\sum_{m=1}^{2n}\frac{1}{m}-
  \frac{1}{2}\sum_{m=1}^n\frac{1}{m}=\psi(2n+1)-\frac{1}{2}\psi(n+1)+\frac{\gamma}{2}.
\end{equation}
Another contiguity relation is
\begin{equation}
K_1(n)-K_0(n) = -\int_0^{\pi/2}\sin\theta\cos\theta\,\sin^22n\theta\,d\theta
 = \frac{4n^2}{4n^2-1}.
\end{equation}
Combining this with (22) we obtain the final result for S-wave states,
\[
\langle(\hbar\kappa/P)\rangle_{n0}=(4/\pi)[\psi(n+1/2)-2n^2/(4n^2-1)+\gamma+\ln 4]
 \qquad{\rm or}
\]
\begin{equation}
\langle\frac{1}{P}\rangle_{n0} = \frac{8an}{h}\left[
  \psi(n+1/2)-\frac{2n^2}{4n^2-1}+\gamma+\ln 4\right],     
\end{equation}
since $h=2\pi\hbar$. For large $n$ this behaves asymptotically as 
\[
\langle(1/P)\rangle_{n0}\sim(8an/h)[\ln(4n)+\gamma-1/2-1/12n^2]+{\rm O}(n^{-3}).
\]

\subsection{The cases $\ell=n-1, n-2$}
The case $\ell=n-1$ is a relatively easy problem because the Gegenbauer 
polynomial collapses to unity and the integral (18) becomes quite trivial.
We have
\begin{eqnarray}
\langle\frac{\hbar\kappa}{P}\rangle_{n\,n-1}&=&\frac{2^{2n-1}n!(n-1)!}{\pi(2n-1)!}
 \int_{-1}^1 (1-x^2)^{n-1}(1+x^2)\,dx \nonumber \\
&=&  \frac{\Gamma(n)\Gamma(n+2)}{\Gamma(n+1/2)\Gamma(n+3/2)}
\end{eqnarray}
or
\begin{equation}
   \langle\frac{1}{P}\rangle_{n\,n-1} = \frac{2\pi a}{h}
   \frac{\Gamma(n+1)\Gamma(n+2)}{\Gamma(n+1/2)\Gamma(n+3/2)}.
\end{equation}
This time the asymptotic behaviour in $n$ is
\[
\langle(1/P\rangle_{n\,n-1}\sim(8\pi a/h)[1 + 3/4n + O(1/n^2)].
\]

In a similar, but somewhat more complicated, vein we can readily treat the case 
$\ell = n-2$ so as to obtain
\[
\langle\frac{\hbar\kappa}{P}\rangle_{n\,n-2} = 
 \frac{(n+2)\Gamma(n-1)\Gamma(n+1)}{\Gamma(n-1/2)\Gamma(n+3/2)}
 \rightarrow 1 + \frac{9}{4n} + O\left(\frac{1}{n^2}\right) .
\]
But, to proceed any further down in $\ell$, a more systematic approach is 
necessary.

\subsection{The general case $0 \leq \ell\leq n-1$}
We can no longer avoid handling the awkward weight occurring in (18).
(Had we been dealing with even powers of $P$ the weight would not have been
too troublesome, by using recurrence properties of the $C_N^\lambda$.) Realising 
that $C_N^\lambda(x)$ is merely an $N$th degree polynomial in $x$, it ought to 
be possible to rewrite it as a combination of $C_M^\mu(x)$ polynomials, with
appropriately chosen $\mu$ (to make the integral (18) more tractable). And 
indeed there is a result which serves this purpose (see \cite{Szego}, eqs 4.10.27
and 4.10.28):
\begin{equation}
\Gamma(\lambda)C_n^\lambda(x) = \sum_{j=0}^{[n/2]}\frac{(n-2j+\mu)}{j!}
   \frac{\Gamma(\lambda-\mu+j)\Gamma(n+\lambda-j)}
        {\Gamma(\lambda-\mu)\Gamma(n+\mu-j+1)}\Gamma(\mu)C_{n-2j}^\mu(x).
\end{equation}
In this connection, note that for integer $j\geq 0$, $\lim_{\epsilon\rightarrow 
0} \Gamma(\epsilon+j)/\Gamma(\epsilon)=\delta_{j0}$, so that (27) reduces 
to a triviality when $\lambda = \mu$.

Let us therefore alter the weight of $\ell$ by 1/2 up and down by expressing
\begin{equation}
C_{n-\ell-1}^{\ell+1}(x) = \sum_{j=0}^{[(n-\ell-1)/2]}
   \beta_{jn\ell}C_{n-\ell-1-2j}^{\ell+1/2}(x)
  = \sum_{j=0}^{[(n-\ell-1)/2]}\gamma_{jn\ell}C_{n-\ell-1-2j}^{\ell+3/2}(x),
\end{equation}
where
\begin{eqnarray}
\beta_{jn\ell}&\equiv&\frac{(n-2j-1/2)}{j!}\frac{\Gamma(\ell+1/2)}{\Gamma(\ell+1)}
 \frac{\Gamma(j+1/2)\Gamma(n-j)}{\Gamma(1/2)\Gamma(n-j+1/2)},\\
\gamma_{jn\ell}&\equiv&\frac{(n-2j+1/2)}{j!}\frac{\Gamma(\ell+3/2)}{\Gamma(\ell+1)}
 \frac{\Gamma(j-1/2)\Gamma(n-j)}{\Gamma(-1/2)\Gamma(n-j+3/2)}.
\end{eqnarray}
Then, using the orthogonality property of the ultraspherical polynomials, we
end up with the {\em single} summations:
\[
\int_{-1}^1\!\!(1-x^2)^\ell[C_{n-\ell-1}^{\ell+1}(x)]^2\,dx\!=\!\!\!\!\!\!
  \sum_{j=0}^{[(n-\ell-1)/2]}\!\!\left( \frac{2^{-\ell}\beta_{jn\ell}}
  {\Gamma(\ell\!+\!1/2)}\right)^{2}\!\!\!
\frac{\pi\Gamma(n\!+\!\ell\!-\!2j)}{(n\!-\!2j\!-\!1/2)(n\!-\!\ell\!-\!2j\!-\!1)!}
\]
and
\[
\int_{-1}^1\!\!\!(1-x^2)^{\ell+1}[C_{n-\ell-1}^{\ell+1}(x)]^2\,dx\!=\!\!\!\!\!\!\!
  \sum_{j=0}^{[(n-\ell-1)/2]}\!\!\left(\!\frac{2^{-1-\ell}\gamma_{jn\ell}}
  {\Gamma(\ell\!+\!3/2)}\!\right)^{2}\!\!\! \frac
  {\pi\Gamma(n\!+\!\ell\!-\!2j\!+\!2)}{(n\!-\!2j\!+\!1/2)(n\!-\!\ell\!-\!2j\!-\!1)!}.
\]
Applying these to (18) we obtain
\begin{eqnarray}
\langle \frac{\hbar\kappa}{P}\rangle_{n\ell} &=& \frac{2n(n-\ell-1)!}{(n+\ell)!}
  \left( \frac{\Gamma(\ell+1)}{\Gamma(\ell+1/2)}\right)^2\,\,
  \sum_{j=0}^{[(n-\ell-1)/2]}\!\! \frac{\Gamma(n+\ell-2j)}{\Gamma(n-\ell-2j)}
   \times \nonumber \\
& & \quad\left[\frac{2\beta_{jn\ell}^2}{n-2j-1/2} - \frac{(n+\ell-2j)(n+\ell+1-2j)
   \gamma_{jn\ell}^2}{(2\ell+1)^2 (n-2j+1/2)} \right].
\end{eqnarray}
Substituting the expressions for $\beta$ and $\gamma$ from (29) and (30)
respectively, we can finally reduce the answer to
\begin{eqnarray}
\langle\frac{\hbar\kappa}{P}\rangle_{n\ell}&=&\frac{2n(n\!-\!\ell\!-\!1)!}{\pi(n+\ell)!}
\sum_{j=0}^{[(n-\ell-1)/2]}\!\!\left(\frac{\Gamma(j\!+\!1/2)\Gamma(n\!-\!j)}
           {\Gamma(j\!+\!1)\Gamma(n\!-\!j\!+\!1/2)}\right)^2 
         \frac{\Gamma(n\!+\!\ell\!-\!2j)}{\Gamma(n\!-\!\ell\!-\!2j)}\nonumber \\
& &\times\left[(2n\!-\!1\!-\!4j)-\frac{(n\!+\!\ell\!-\!2j)(n\!+\!1/2\!-\!2j)
   (n\!+\!\ell\!+\!1\!-\!2j)}{(2j-1)^2(2n-2j+1)^2}\right].
\end{eqnarray}
We have not succeeded in simplifying this any further, except for the earlier
simple cases. In Table 1, we have provided the first few expectation
values of $2\pi\hbar\kappa/P$ (for $n$ and $\ell$ up to 6) as derived from (32).
Remember that $\kappa = 1/na$, where $a$ is the Bohr radius. 

\begin{table}[tbp]
\begin{center}
{\normalsize
\begin{tabular}{|c|cccccc|}
\hline
$\ell\backslash n$ & 1 & 2 & 3 & 4 & 5 & 6\\ \hline \hline
 0 & $\frac{32}{3}$ & $\frac{256}{15}$ & $\frac{2144}{105}$ & $\frac{1024}{45}$ & 
  $\frac{85088}{3465}$ & $\frac{1172224}{45045}$ \\\hline
 1 &       -       & $\frac{128}{15}$ & $\frac{256}{21}$ & $\frac{512}{35}$ & 
  $\frac{57088}{3465}$ & $\frac{809344}{45045}$ \\\hline
 2 & - & - & $\frac{4096}{525}$ & $\frac{16384}{1575}$ & $\frac{299088}{24255}$ &
  $\frac{21856256}{1576575}$ \\\hline
 3 & - & - & - & $\frac{16384}{2205}$ & $\frac{32768}{3465}$ &
  $\frac{950272}{85995}$ \\\hline
 4 & - & - & - & - & $\frac{524288}{72765}$ & $\frac{8388608}{945945}$ \\\hline
 5 & - & - & - & - & - & $\frac{2097152}{297297}$ \\\hline
\end{tabular}}
\end{center}
\caption{ Calculated values of $\langle 2\pi\hbar\kappa/P \rangle_{n\ell}$ for principal
quantum number $n$ running from 1 to 6, as given by (32).}
\end{table}

\section{Sum rules and integral representation}

There exists an interesting set of sum rules for 
$\langle\hbar\kappa/P\rangle_{n\ell}$ which hail from a particular
form of the addition theorem for Gegenbauer polynomials (see \cite{Magnus},
\S 5.3), and which produce a neat integral representation. Start with
the particular 4-dimensional case
\begin{eqnarray*}
C_{n-1}^1(\cos^2\theta+\sin^2\theta\cos\psi)&=&\sum_{\ell=0}^{n-1}
 \frac{(2\ell+1)(\ell!)^2\Gamma(n-\ell)}{\Gamma(n+\ell+1)}
 (2\sin\theta)^{2\ell}\times \\
& & \qquad  [C_{n-\ell-1}^{\ell+1}(\cos\theta)]^2P_\ell(\cos\psi).
\end{eqnarray*}
On putting $\cos\theta=x$, we find that the rhs has a close connection with
the rhs of eq (18) and so recognize that
\begin{equation}
\int_{-1}^1 (1+x)^2\,C_{n-1}^1(x^2+(1-x^2)\cos\psi)\,dx = 
\frac{\pi}{2n}\sum_{\ell=0}^{n-1} (2\ell+1)P_\ell(\cos\psi)
\langle\hbar\kappa/P\rangle_{n\ell}.
\end{equation}
Since $C_{n-1}^1(y)=U_{n-1}(y)$, we can simplify this result to
\begin{equation}
\sum_{\ell=0}^{n-1}(2\ell+1)P_\ell(y)\langle\frac{\hbar\kappa}{P}\rangle_{n\ell}
=\frac{2n}{\pi}\int_{-1}^1(1+x^2)U_{n-1}(x^2+(1-x^2)y)\,dx.
\end{equation}
To obtain the sum rules, simply set $y=1$ and $y=-1$:
\begin{equation}
\sum_{\ell=0}^{n-1}(2\ell+1)\langle\frac{\hbar\kappa}{P}\rangle_{n\ell} = 
\frac{2n^2}{\pi}\int_{-1}^1(1+x^2)\,dx = \frac{16n^2}{3\pi},
\end{equation}
\begin{equation}
\sum_{\ell=0}^{n-1}(2\ell+1)(-1)^\ell\langle\frac{\hbar\kappa}{P}\rangle_{n\ell}
= \frac{2n}{\pi}\int_{-1}^1(1+x^2)U_{n-1}(2x^2-1)\,dx.
\end{equation}
In order to evaluate the rhs of eq (36) first define 
$J_n=\int_{-1}^1 U_n(2x^2-1)\,dx$. The rhs of (36) can then be rewritten as 
$(J_n+6J_{n-1}+J_{n-2})/4$, so it only remains to determine $J_n$. This can be done
via an easily established recurrence relation, namely $J_n+J_{n-1} = 2/(2n+1)$.
Hence we deduce that
\[ 2J_n=\psi(n/2+5/4)-\psi(n/2+3/4)+(-1)^n\pi. \]
Therefore
\begin{equation}
\sum_{\ell=0}^{n-1}(2\ell+1)(-1)^\ell\langle\frac{\hbar\kappa}{P}\rangle_{n\ell}
= \frac{n}{\pi}\left[\psi(n+\frac{3}{4})-\psi(n+\frac{1}{4})+\frac{4n}{4n^2-1}
   +(-1)^{n-1}\pi   \right].
\end{equation}

Finally, on multiplying (34) by a Legendre polynomial in $y$ and integrating
over (-1,1), we obtain a neat double integral representation for the 
expectation value. Thus
\begin{equation}
\langle\frac{\hbar\kappa}{P}\rangle€_{n\ell} =\frac{n}{\pi}\int_{-1}^1
\,(1+x^2)\left[\int_{-1}^1 P_\ell(y)U_{n-1}(x^2+(1-x^2)y)\,dy\right]dx.
\end{equation}
One may verify in particular cases that this produces the same results as 
the series (32) but, more significantly, it allows us to obtain a series
representation of $\langle \hbar\kappa/P\rangle_{n\ell}$, which is different
from (32).

To arrive at this new result, we make use of the fact (see \cite{Magnus}, 
\S 5.7.2) that
$$
 U_{n-1}(z)=\sqrt{\pi}\sum_{j=0}^{n-1}
 \frac{(-1)^j (n+j)!(1-z)^j}{j!(n-j-1)!2^{j+1}\Gamma(j+3/2)}, 
$$
and notice the factorization, $1-(x^2+(1-x^2)y)=(1-x^2)(1-y)$. Thus (38)
factorizes into two parts. The integral over $x$ is easily done, but
the integral over $y$ is a bit harder (requiring Rodrigues' formula and
an integration by parts).
Carrying out the necessary manoeuvres we end up with
\begin{equation}
\langle\frac{\hbar\kappa}{P}\rangle_{n\ell}=\sum_{j=0}^{n-\ell-1}
\frac{(-1)^j\, n\,(\ell+j+2)(n+\ell+j)![(\ell+j)!]^2}
     {(n-\ell-j-1)!(2\ell+j+1)!j!\Gamma(\ell+j+3/2)\Gamma(\ell+j+5/2)}.
\end{equation}
This is more compact than the series (32).

\section{Asymptotic behaviour}
There are 3 regimes to consider for large $n$ which we shall
look at in turn.

\subsection{$\ell/n \ll 1$}
Returning to eq (18) we may replace $\Gamma(n-\ell)/\Gamma(n+\ell+1)$ 
asymptotically by $n^{-2\ell -1}$ and use the approximation
(see \cite{Magnus}, \S 5.3.3),
\[
C_{n-\ell-1}^{\ell+1}(\cos 2\theta)\sim \frac{n^\ell}{2^\ell \ell!}
 \frac{\cos(2n\theta-\pi(\ell+1)/2)}{(\sin\,2\theta)^{\ell+1}},\qquad n\gg 1,
\,0<\theta <\pi/2.
\]
This tells us that
\[
\langle\frac{\hbar\kappa}{P}\rangle_{n\ell}\sim \lim_{\epsilon\rightarrow 0+}\frac{16}{\pi}
 \int_{\epsilon}^{\pi/2-\epsilon}
\frac{\cos^4\theta \cos^2(2n\theta-\pi(\ell+1)/2)}{\sin 2\theta} d\theta,
\]
assuming the integral exists. Hence we obtain the asymptotic difference
between two $\ell$ values differing by two:
\[
\langle\frac{\hbar\kappa}{P}\rangle_{n\,\ell+1}-\langle\frac{\hbar\kappa}{P}
\rangle_{n\,\ell-1}\sim\lim_{\epsilon\rightarrow 0+}\frac{16}{\pi} 
\int_{\epsilon}^{\pi/2-\epsilon}d\theta \,\,\frac{\cos^4\theta}{\sin 2\theta}\times
\]
\[\qquad\qquad\qquad
\left[\cos^2(2n\theta-\pi(\ell+2)/2)-\cos^2(2n\theta-\pi\ell/2)\right]=0.
\]
But we have already shown in (24) that $\langle\hbar\kappa/P\rangle_{n\ell}
\sim 4\psi(n+1/2)/\pi\,$ for $\ell=0$, from which we conclude that
\begin{equation}
\langle\frac{\hbar\kappa}{P}\rangle_{n\ell} \sim 4\psi(n+1/2)/\pi
\end{equation}
for $n\gg 1$ and modest values of $\,\ell$. This result diverges 
logarithmically like $(4\log n)/\pi$ as $n\rightarrow\infty$.

\subsection{$\ell/n$ near 1}
In this case return to the series (32) and put $n=\ell+1+\delta\,$ where 
$\delta/n\ll 1$. Making use of the asymptotic approximation,
\begin{equation}
\lim_{z\rightarrow\infty}\frac{\Gamma(z+a)}{\Gamma(z+b)} \sim
 z^{a-b}\left[1 + \frac{1}{2z}(a-b)(a+b+1)+O(\frac{1}{z^2})\right],
\end{equation}
the sum collapses to
\[
\langle\frac{\hbar\kappa}{P}\rangle_{n\,n-1-\delta}\sim\frac{1}{\pi}
\sum_{j=0}^{[\delta/2]}(2n)^{-2j}\frac{\Gamma(1\!+\!\delta)}
 {\Gamma(1\!+\!\delta\!-\!2j)}\!\!\left(\!\!\frac{\Gamma(j\!+\!1/2)}{\Gamma(j\!+\!1)}
 \!\right)^2\!\!\left[2\!-\!\frac{1}{(2j\!\!-\!\!1)^2}+O(\frac{1}{n})\right].
\]
Thus $j=0$ is the dominant term in the expansion; this is multiplied by a
subdominant factor of order $1/n$ via (32), producing the final estimate,
\begin{equation}
\langle\frac{\hbar\kappa}{P}\rangle_{n\, n-1-\delta}\sim 1+\frac{3(2\delta+1)}{4n}
 + O(\frac{1}{n^2}).
\end{equation}
This agrees with the results of section 3.2 where the cases
$\delta=0$ and 1 were studied.

\subsection{$\ell/n$ is finite}
This regime is the trickiest to deal with, interpolating between (40) and (42)
as it does. Although we have not succeeded in obtaining the analytical dependence
on $\lambda\equiv \ell/(n-1)$, one may readily establish numerically that the sum
(39) tends to a constant in the limit of large $n$ and increases as $\lambda$ approaches
zero; for instance when $\lambda=1/2$ the value is 1.975; when $\lambda=1/4$, 
the value is 2.88; when $\lambda=1/ 8$, the value is 3.77, etc. This dependence
would appear to be logarithmic as is indicated by (24).

\section{Conclusions}
In this paper we have principally concentrated on the evaluation of 
$\langle 1/P \rangle$ --- a challenging problem --- because 
it has special significance for Born reciprocity, but the methods we have used 
can no doubt be extended to general functions of momentum via eqs (16) and (17).
In fact such calculations have been carried out in \cite{AYGD} for
expectation values of $P^N$ and $\log P$ to which the reader is referred. At 
any rate, the conclusion of our analysis is that the $1/p$ perturbation has an 
increasingly disruptive effect on the higher $n$ states having the lowest 
$\ell$ (the dependence is logarithmic in $n$). This hails from the 
full effective potential \cite{DL} which shows a maximum below zero energy,
at $E_0 = bL -2\alpha\sqrt{bL}$, where $L$ signifies the classical 
angular momentum.

\end{document}